\documentclass[useAMS,usenatbib,usegraphicx]{mn2e}

\title[Multiplicity of high-mass stars]
{A spectroscopic survey on the multiplicity of high-mass stars}
\author[R.~Chini et al.]
{R.~Chini$^{1,2}$,
 V.H.~Hoffmeister$^{1}$,
 A.~Nasseri$^{1}$,
 O.~Stahl$^{3}$,
 and H.~Zinnecker$^{4}$\\
$^{1}$Astronomisches Institut, Ruhr-Universit\"at Bochum, Universit\"atsstra\ss{}e 150, D-44780 Bochum, Germany\\
$^{2}$Instituto de Astronom\'{i}a, Universidad Cat\'{o}lica del Norte, Antofagasta, Chile\\
$^{3}$ZAH, Landessternwarte Heidelberg-K\"onigstuhl, D-69117 Heidelberg, Germany\\
$^{4}$SOFIA Science Center, Mailstop 211-3, Moffett Field, CA94035, USA}
\begin{document}

\date{Accepted 20xx . Received 2011 June }

\pagerange{\pageref{firstpage}--\pageref{lastpage}} \pubyear{2011}

\maketitle

\label{firstpage}

\begin{abstract}
The formation of stars above about twenty solar masses and their apparently high multiplicity remain heavily debated subjects in astrophysics. We have performed a vast high-resolution radial velocity spectroscopic survey of about 250 O- and 540 B-type stars in the southern Milky Way which indicates that the majority of stars ($> 82\%$) with masses above 16 solar masses form close binary systems while this fraction rapidly drops to 20\% for stars of 3 solar masses. The binary fractions of O-type stars among different environment classes are: clusters ($72 \pm 13\%$), associations ($73 \pm 8\%$), field ($43 \pm 13\%$), and runaways ($69 \pm 11\%$). The high frequency of close pairs with components of similar mass argues in favour of a multiplicity originating from the formation process rather than from a tidal capture in a dense cluster. The high binary frequency of runaway O stars that we found in our survey (69\% compared to $19 - 26\%$ in previous surveys) points to the importance of ejection from young star clusters and thus supports the competitive accretion scenario.

\end{abstract}


\section{Introduction}

The most massive stars, historically classified as O-type stars, are rare objects with masses above 16\,M$_\odot$, typically found at large distances from the sun; early B stars ($M > 8\,$M$_\odot$) also contribute to the group of high-mass objects. The formation of these O- and B-type stars is still under debate and can be explained by various models \citep{ZY07}. Both observations of massive circumstellar disks \citep{SCK01,CHKetal04,CNO05,PCS05,KHMetal10} and theoretical calculations \citep{YS02,KKMetal09,KKBetal10} seem to favour the accretion scenario. However, the high multiplicity among high-mass stars \citep{PBHetal99,MHGetal09} might alternatively support a merging process of intermediate-mass stars \citep{BBZ98}.

Massive binary stars are believed to be the progenitors of a variety of astrophysical phenomena, e.g. short gamma-ray bursts \citep{ELPetal89,P91,NPP92}, X-ray binaries \citep{M08}, millisecond pulsar systems and double neutron stars \citep{VDH07}. Even more relevant, the multiplicity of stellar systems is a crucial constraint for the various star formation scenarios, particularly if the multiplicity fraction were a function of stellar mass. While many massive stars are found to be part of binary or multiple systems, comprehensive statistics on close binaries is still missing. The smallest separations are expected to be around 0.2 AU, resulting in orbital periods of a few days only. Below this minimum distance the binary system will merge to form a single object.

The vast parameter space in possible orbital periods and mass ratios requires different, partly overlapping complementary methods that have their own limitations and their observational biases \citep{SE11}: High-resolution imaging like adaptive optics or interferometric techniques serve for systems with wider separations and mass ratios $q = M_2 / M_1$ between 0.01 and 1 while high-resolution spectroscopy is biased toward finding close companions and those that are a significant fraction of the primary's mass ($q > 0.1$). We note that the inclination of the orbit with respect to the observer and the eccentricity are other crucial limitations for the spectroscopic detection of close multiple systems.

An adaptive optics survey of about a third of the known galactic O stars revealed visual companions in 27\% of the cases \citep{SE11}. Speckle interferometry for most of the galactic O stars showed companions for 11\% of the sample \citep{MHGetal09}. Moreover, the same study claimed that 51\% of the O-type objects are spectroscopic binaries (SBs), based on an extensive review of the literature. This is in accord with a recent spectroscopic survey finding that among $\sim 240$ southern galactic O and WN stars more than one hundred stars show radial velocity ($RV$) variations larger than 10\,km/s \citep{BGAetal10}. Recently, a high-resolution imaging campaign of 138 fields containing at least one high-mass star yielded a multiplicity fraction close to 50\% \citep{M10}. In summary, the spectroscopic binary frequency of high-mass stars so far observed and reported in the literature is moderately high, while the visual binary fraction is low. Little or no discussion as to why the multiplicity for O-type stars is high has been offered up to now nor have conclusions been drawn so far about which of the competing high-mass star formation models best explains the hitherto known trends in the stars' multiplicity.

There have been several surveys of B star duplicity in the past. In a sample of 109 B2 - B5 stars there were 32 (29\%) spectroscopic and 49 (45\%) visual binaries yielding a total binary frequency of 74\% \citep{AGL90}. A spectroscopic survey of 83 late B-type stars revealed that 24\% of the stars had companions with mass ratios greater than 0.1 and orbital periods less than 100 days \citep{W78}. A speckle interferometry survey of the Bright Star Catalogue resolved 34 of 245 B stars into binaries corresponding to a multiplicity fraction of ~14\% \citep{MHHetal87,MHHetal93}.

Another speckle interferometry survey of 48 Be stars revealed a similar binary fraction of $10\% \pm 4\%$ \citep{MTGetal97}. A further comparison study, based on adaptive optics IR imaging and probing separations from 20 to 1000 AU for 40 B and 39 Be stars derived the same qualitative result, i.e. that the multiplicity of B and Be stars are identical. This time, however, the binary fractions were $29\% \pm 8\%$ for the B stars and $30\% \pm 8\%$ for the Be stars \citep{OP10}. Finally an adaptive optics photometry and astrometry survey of 70 B stars revealed 16 resolved companions (23\%) \citep{RTT07}. In summary, the overall multiplicity fraction of high-mass stars seems to decrease with stellar mass.

In the present paper we investigate the multiplicity fraction in the stellar mass range of about $3 - 80\,M_\odot$ and for mass ratios $q > 0.2$.

\section{Observations}

\begin{figure}
	\includegraphics*[width=84mm]{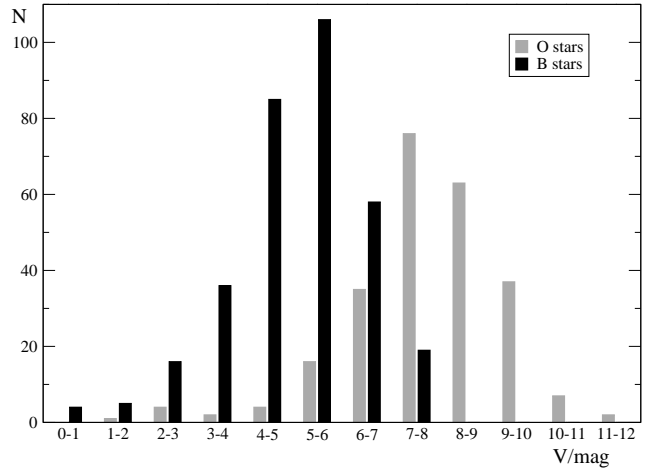}
	\caption{
    		Distribution of $V$ magnitudes for the samples of O-type stars (grey) and B-type
            stars (black) columns. The B stars are on average brighter which results in better $S/N$ spectra and thus increases the chance for detecting fainter companions.
			}
\label{histo_mag}
\end{figure}

We have performed a comprehensive spectroscopic survey on a large representative sample of 249 O- and 581 B-type stars. The O-stars were taken from the Galactic O-Star Catalogue V.2.0 \citep{SMWetal08} (GOSC) which is assumed to be complete to $V \le 8$\,mag. The spectral classification in the GOSC is currently under revision: A new version has been presented by \citep{SMWetal11} for 184 stars with $B \le 8$ and north of $\delta = -20^\circ$; a second revision is in preparation by the same authors. The B stars were selected from the HIPPARCOS archive in a way that each subclass (B0 - B9) contains roughly an equal amount of stars; 50\% of the B stars form a volume-limited sample with $d < 125$\,pc. The distribution of visual magnitudes is displayed in Fig.~\ref{histo_mag} showing that the B-star sample contains on average brighter stars than the O-star sample.

Using the high-resolution spectrograph BESO \citep{FCH11} at the Hexapod Telescope at the Universit\"atssternwarte Bochum near Cerro Armazones in Chile we obtained 3632 multi-epoch optical spectra were obtained with BESO. The observing period started in January 2009 and is still going on. The spectra comprise a wavelength range from 3620 to 8530\,\AA\, and provide a mean spectral resolution of $R = 50.000$. The entrance aperture of the star fibre is $3\farcs4$. The integration time per spectrum was adapted to the published visual brightness of each star. It was our primary goal to monitor a large number of stars rather than to obtain a very high $S/N$ for individual stars. As a consequence of this strategy potential companions fainter than $\Delta V \sim 2$\,mag are barely visible in our spectra, thus decreasing the chance for detecting SB2s. Converting this brightness difference of 2\,mag into a mass difference we are sensitive to mass ratios $q > (0.18 -  0.40)$ for O5 - O9 stars and $q > (0.43 - 0.55)$ for B stars. In other words, the detectable companions of an O5 star range from O5 to about B2, those of a B9 star from B9 to about A7.

Additional spectra were collected during various observing runs with FEROS at the ESO 1.5\,m and the MPG/ESO 2.2\,m telescopes, both located on La Silla, Chile. Because BESO is a clone of FEROS the instrumental parameters are identical to those described above. These spectra cover a time span between 2006 and 2008.

Finally we extracted 1009 O-star spectra and 133 B-star spectra from the ESO archive covering an observing period from 1999 to May 2010. These spectra were also obtained with FEROS and processed in the same way as the BESO spectra.

Until today there are 2565 O- and 2209 B-star spectra in our archive. On average we collected ten spectra for each O-star with a minimum of two spectra for those objects which were known to be multiple systems before and a maximum of 30 spectra in such cases where multiplicity was not immediately obvious. The spectra are separated by days, weeks and months. So far we have observed 540 stars from our B-type sample with an average number of four spectra per star. We stopped temporarily collecting data for those B-stars where the multiplicity became obvious during the first two spectra.

\begin{figure}
	\includegraphics*[width=84mm]{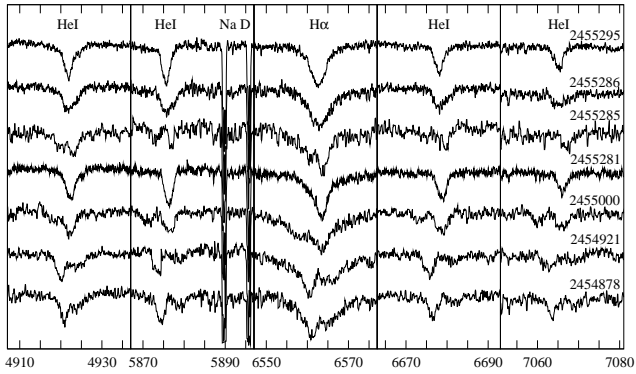}
	\caption{
			Multi-epoch BESO spectra obtained between JD 2454878 and 2455295
            for the double-lined O star binary HD\,104649. The selected spectral regions show several He\,I lines, H$\alpha$ and the interstellar NaD lines. The change from a single- to a double-lined structure is clearly visible.
			}
\label{HD104649}
\end{figure}

All data were reduced with a pipeline based on the MIDAS package developed for the ESO FEROS spectrograph. A quantitative analysis was done with standard IRAF line-fit routines allowing the detection of line shifts in single-line spectroscopic binaries (SB1) or line deformation and/or separations in double-line spectroscopic binaries (SB2). For the O- and early B-stars we used exclusively He lines for identification; He\,I ($\lambda 5875$\,\AA) was particularly useful due to the nearby interstellar Na\,I doublet ($\lambda\lambda  5890, 5896$\,\AA) allowing for a sensitive verification of any relative line shift. For later B-types we had to rely primarily on hydrogen lines. A typical set of multi-epoch spectra is shown in Fig.~\ref{HD104649}.

\section{Results}

The data of our survey will be published in electronic format. Table~\ref{t:example} shows the corresponding entries: Name, Sp.T. = spectral type, No. = number of spectra, and Mult. = multiplicity status (C = constant \emph{RV}, SB1, SB2, U = unknown) and some example results.

\begin{table}
\caption{Spectroscopic results for 248 O- and 581 B-type stars.}
\begin{tabular}
{llcc}
\hline
Name        & Sp.T. & No.   & Mult.\\
\hline
HD 000560   & B9\,V     &  10   &   C  \\
HD 014951   & B7\,IV    &  10   &  SB2 \\
HD 017543   & B2\,IV    &  10   &  SB1 \\
HD 035588   & B2.5\,V   &   6   &   U  \\
\hline
\end{tabular}
\label{t:example}
\end{table}

\begin{figure}
	\includegraphics*[width=84mm]{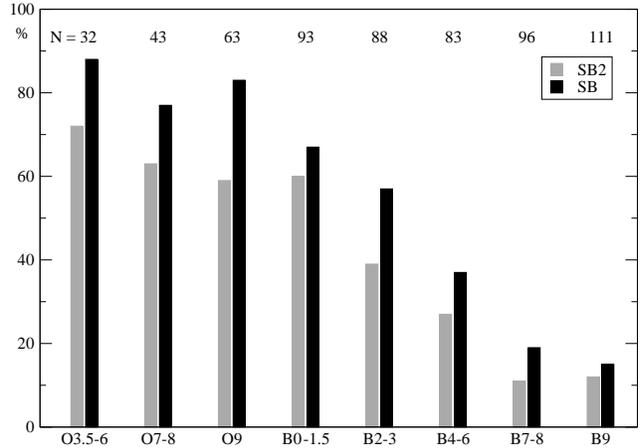}
	\caption{
			Multiplicity vs. spectral type of O- and B-type stars. The black columns
            denote the total percentage of spectroscopic binaries, the grey columns display the percentage of double-lined spectroscopic binaries. The number \emph{N} on top of the columns denote the number of objects in each spectral bin.
			}
\label{histo_mult}
\end{figure}

We detect \emph{RV} variations for about 82\% of the O stars brighter than $V = 8$\,mag while for the B stars the variability fraction decreases from 61\% (B0) to $\sim 15\%$ (B9) (Fig.~\ref{histo_mult}). For fainter O stars the cumulative percentage of $RV$ variations diminishes to 71\% ($V < 9$\,mag) and 68\% ($V < 10$\,mag) reflecting the decreasing instrumental sensitivity. In general, our observed multiplicities are lower limits due to the unknown orbit inclination and the constraints for the detectable minimum mass ratio.

The overall percentage of SBs for O stars is higher than found before in similar investigations and is due to the numerous spectra obtained for a single star, which reveal mainly variations within days and weeks that were not obvious in previous studies. As mentioned above \citet{MHGetal09} report a general spectroscopic multiplicity fraction of 51\% for the population of Galactic O stars. Individual nearby clusters were found to have binary fractions between 0\% and 63\% with an average value of 44\% (see e.g. \citet{SE11} for an overview).

Inspecting those 60 O stars that did not show any $RV$ variations throughout the last six years, we find that additional 13 out of 29 stars observed through adaptive optics measurements \citep{MTGetal97} or speckle interferometry \citep{MHHetal93} possess visual companions. These complementary data increase the total percentage of multiple systems for stars brighter than $V = 8$\,mag to 91\% leading us to suggest that basically all O-stars are members of multiple systems. This finding also goes beyond the results summarized by \citet{SE11} who obtain a total minimum multiplicity fraction close to 70\%.

Among the SBs in the present study 82\% of the spectra for O stars with $V \le 10$\,mag contain more or less separated multiple lines (SB2) reflecting that the majority of systems contains pairs of similar mass. This is in agreement with results by \citet{KF07} who found that massive stars preferentially have massive companions. From our current time coverage, however, it is not yet possible to derive orbital periods and constrain semi-amplitudes of radial velocities and hence accurate binary component mass ratios.

Due to the limited spectroscopic binary studies for B stars a comparison between our work and previous investigations is difficult.\ \citet{KBZetal05} studied the binary population in the Scorpius OB2 association. Summarizing the results from all techniques they find multiplicity fractions of $\sim 80\%$ and $\sim 50\%$ for B0 - B3 and B4 - B9, respectively. While the numbers for the early B-types are compatible with our study the multiplicity rate for the late B-types is slightly higher in Sco OB2; the latter difference might be due to the different observing techniques, because without the adaptive optics data of \citet{KBZetal05} the multiplicity rate would drop by about 10\% for the late B-types. \citet{KBPZetal07} repeated the analysis of the primordial binary population in Sco OB2: one set of spectroscopic data in this study comes from \citet{Letal87} who found that the binary fraction is at least 30\% for all early-type stars but might be as high as 74\% if all reported \emph{RV} variations were due to binaries. Another spectroscopic study \citep{BV97} included in the analysis by \citet{KBPZetal07} yields a similar range between 28\% and 76\%. However, the dependence on stellar mass has not been addressed in these studies.

\citet{WOG10} investigated the binarity of 25 Herbig Be stars with spectro-astrometry and derive a high binary fraction of 74\%. This result, however, is in contrast to the AO study of \citet{OP10} who found a binary fraction of $30\% \pm 8\%$ for 39 Be stars. The same authors claim that the binarity of normal B stars is $29\% \pm 8\%$ and thus identical to those of Be stars. In summary, the multiplicity of B stars is still a subject that needs further investigation; the existing results -- including ours -- likely suffer from various biases. We actually might expect to observe more binary stars amongst late type B stars compared to O stars, however, then their mass ratios must be larger than what is covered by our study.

\begin{figure}
	\includegraphics*[width=84mm]{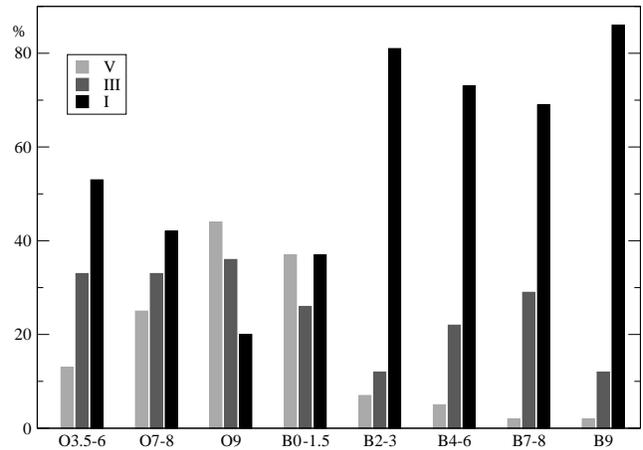}
	\caption{
			Luminosity classes for each spectral bin.
			}
\label{histo_evo}
\end{figure}

One may speculate whether the decrease of the binarity fraction between O- and late B-type stars is due to an \emph{evolutionary} effect: an O-type star -- whether on the main sequence or not -- is much younger than a late B-type star on the main sequence and is likely more evolved due to its rapid hydrogen burning. If evolution "destroys" binarity then evolved stars (i.e. luminosity classes III and I) should have lower binarity fractions than stars on the main sequence.  Fig.~\ref{histo_evo} shows the distribution of luminosity classes I, III, and V for the different spectral type bins in our sample. Obviously those spectral types with the highest binarity fraction (O3 -- B1) have the highest fraction of evolved stars while late B-Type stars with the lowest binarity fraction comprise only few evolved stars. This excludes the possibility that evolution influences binarity. On the other hand, one can exclude a pure \emph{age} effect as SB2 binaries are all very tight and hard to break up. They could dynamically interact in dense cluster cores and get ejected, but breaking up such "hard" SB2s is not very likely.

The multiplicity of high-mass stars seems to depend on their environment. It was found that the binary frequency among O stars in clusters and associations is much higher than among field stars (which have no apparent nearby cluster) and runaway stars (O stars with peculiar $RV$s in excess of 40\,km/s or remote from the Galactic plane). The spectroscopic binary fraction in clusters and associations, in the field, and among runaways obtained in two different studies \citep{G87,MGH98} were 55 (61)\%, 45 (50)\% and 19 (26)\%, respectively. The O stars in the present sample can be assigned to clusters (44), associations (123), field (25), and runaways (54). For the classification we have used the designations in the GOSC \citep{SMWetal08} refined by a recent investigation which turned 73 out of 93 "field stars" from the GOSC into runaways originating from open clusters \citep{SR08}. In those cases where stars in clusters (4), associations (3) or in the field (47) were classified as runaway stars we have adopted this new designation. The fractions of SBs within the individual groups are: clusters ($72 \pm 13\%$), associations ($73 \pm 8\%$), field ($43 \pm 13\%$), and runaways ($69 \pm 11\%$); the errors are based on the number statistics. While our results for the clusters/associations and the field are compatible with previous studies \citep{G87,MGH98} our binary fraction among the runaways ($69 \pm 11\%$) is significantly higher compared to \citet{G87} ($19 \pm 5\%$) and \citet{MGH98} ($26 \pm 5\%$), respectively. If field stars are dynamically ejected from a cluster, one expects an equal fraction of singles and binaries expelled in opposite directions \citep{HBZ00,HBZ01,GPZ04}.

\section{Discussion}

While the direct observation of disks around the earliest O stars will remain a fundamental challenge to test model computations for individual high-mass systems, binary star statistics offers other constraints on hydrodynamical simulations of star forming clusters. Early calculations of the fragmentation of an isothermally collapsing cloud have already shown that binary systems and hierarchical multiple systems are frequently obtained \citep{L78}. Although the variety of processes and their sequences are manyfold \citep{BBB02} there is consensus that the accretion processes favor the formation of binaries with mass ratios $q \sim 1$ for those systems with separations below 10\,AU, (e.g. \citet{C07} and references therein).

Nevertheless, the simulations almost never produce binaries with $q < 0.5$. This appears to be a general problem of turbulent fragmentation calculations which seems to origin from gas accretion onto a proto-binary \citep{GWW04}. Independent of how the binary system was formed, the infalling material has a high specific angular momentum compared to the binary and thus will preferentially accrete onto the secondary  \citep{BB97}. \citet{B00} even predicts that binaries will evolve rapidly towards $q \sim 1$ regardless of their initial mass ratio. For the special case of massive stars recent theoretical calculations claim that close high-mass stellar twins can in principle be formed via fragmentation of a disk around a massive protostar and subsequent mass transfer in such close, rapidly accreting oversized proto-binaries \citep{KT07}. This scenario provides a natural explanation for the numerous high-mass spectroscopic binaries with mass ratios close to unity.

The high binary frequency for clusters found in the present study exclude random pairing from a classical IMF as a process to describe the similar-mass companions in massive binaries. Our results make binary-binary interactions inside clusters very probable and thus can explain the high binary fraction among runaway stars. Because ejection requires a cluster origin of the binaries \citep{K01} it supports the model of competitive accretion within the cluster environment. Finally, the small number of field O stars - only six "certified" stars remain currently in this category - suggests that probably all O stars are born in clusters.

Radiation-hydrodynamic simulations show that, during the collapse of a massive prestellar core, gravitational and Rayleigh-Taylor instabilities channel gas onto the star system through non-axisymmetric disks. Gravitational instabilities lead to a fragmentation of the disk around the primary star and form a massive companion. Radiation pressure does not limit stellar masses, but the instabilities that allow accretion to continue lead to small multiple systems \citep{KKMetal09}.

The current study was a pure discovery project that aimed exclusively at the multiplicity statistics in the mass range $3 < M \rm{[}M_\odot\rm{]} \simeq 80$. In a next step we will study the orbital properties and the spectral types of the individual components to obtain a statistically relevant archive for high-mass binary systems. Likewise, we are currently performing a photometric monitoring program to search for eclipsing binaries in order to determine accurate masses for O stars. From the fact that most O-type stars occur as similar mass binaries we conclude that their absolute magnitude calibration has been overestimated; in the worst case the tabulated O star magnitudes are wrong by 0.75\,mag at each wavelength.

\section*{Acknowledgments}

We wish to thank K. Fuhrmann, M. Haas, B. Reipurth and B. Stecklum for helpful discussions. This publication is supported as a project of the  Nordrhein-Westf\"alische Akademie der Wissenschaften und der K\"unste in the framework of the  academy program by the Federal Republic of Germany and the state Nordrhein-Westfalen.

\bsp

\label{lastpage}

\end{document}